\documentclass[%
 reprint,
nofootinbib,
 amsmath,amssymb,
 aps,
prd
]{revtex4-2}

\usepackage{graphicx}
\graphicspath{{img/}}
\usepackage{dcolumn}
\usepackage{bm}
\usepackage{hyperref}
\usepackage{cleveref}
\usepackage[separate-uncertainty=true]{siunitx}
\usepackage{caption}
\usepackage{subcaption}
\usepackage{placeins}
\usepackage{amsmath}

\newcommand{\componentlibrary}{{This sketch was created using the \href{http://www.gwoptics.org/ComponentLibrary/}{Component Library by Alexander Franzen}, licensed under \href{https://creativecommons.org/licenses/by-nc/3.0/}{CC BY-NC 3.0}.}}

\begin{document}

\title{Scattered light reduction in Sagnac Speed Meters with Tunable Coherence}

\author{Leonie Eggers}
 \email{leonie.eggers@uni-hamburg.de}

\author{Daniel Voigt}%
 \email{daniel.voigt@uni-hamburg.de}
\author{Oliver Gerberding}
 \email{oliver.gerberding@uni-hamburg.de}

\affiliation{Institute of Experimental Physics, 
University of Hamburg, 
Luruper Chaussee 149, 
22761 Hamburg
}

\date{\today}

\begin{abstract}
Sagnac Speed Meter and ring resonators can be used as high precision instruments, but they are limited in their sensitivity through scattered light causing non-linear noise. Here, we experimentally demonstrate a technique called Tunable Coherence, where the long coherence length of the laser is broken in a controlled way, to suppress the coupling of scattered light in a Sagnac interferometer. We demonstrate a scattered light suppression of $\SI{24.2}{\decibel}$ in a Sagnac interferometer and discuss the experimental limitations. Further, we show an analytical discussion on how Tunable Coherence could be a fundamental solution to light scattering back from optical surfaces into the counter propagating beam, which is an issue particularly in ring resonators.
\end{abstract}

\maketitle

\section{Introduction}  
Laser interferometers are used in a wide range of applications, one being ground-based gravitational wave detectors, which currently employ an enhanced Michelson interferometer layout. Since the first detection in 2015, now known as GW150914, by the advanced LIGO detectors~\cite{lvk2016}, many more detections provided further
insights in astrophysical and cosmological phenomena~\cite{lvk2016, lvk2017, abbott2021, Kasen2017}. While the network of current detectors~\cite{theligoscientificcollaboration2017} consisting of the Advanced LIGO, Advanced Virgo~\cite{acernese2015}, KAGRA~\cite{akutsu2021} and GEO600~\cite{luck2010} are further improved and upgraded to increase sensitivities, the next generation of observatories like the Einstein Telescope~\cite{punturo2010} and Cosmic Explorer~\cite{reitze2019} are already being planned and will push the limits even further.

For these future observatories, Sagnac Speed Meters are a possible alternative topology~\cite{chen2003,purdue2002,wang2013} as they offer advantages for quantum noise reduction due to speed being a quantum non-demolition observable~\cite{chen2003}. However, the Sagnac topology suffers in the same way from scattered light noise as the currently implemented Michelson topologies. Additionally, as it features two counter-propagating beams, another effect of light scattering from one propagation direction into the other, referred to as  backscatter~\cite{zhang2019,pascucci2018}, introduces even more phase noise and a power imbalance. This is especially problematic in the associated ring resonators~\cite{giovinetti2024,liu2020,divirgilio2024,maccioni2022} and also impacts other applications of Sagnac type topologies like gyroscopes~\cite{ringlaser_igel2021,ringlaser_Korth2015,ringlaser_schreiber2023}. While here the impact of backscatter is currently reduced through post-processing techniques, no general solution exists~\cite{hurst2014,petrukhin2021}.

Both backscatter and scattered light noise common to Sagnac and Michelson topologies are limiting factors for sensitivity, especially in the low-frequency regime. The latter is caused by light leaving the main interferometer beam in unintended ways before coupling back into the readout, picking up an additional path length and time-dependent phase changes~\cite{vinet1997,ottaway2012}. Current gravitational wave detectors already employ different methods to reduce scattered light in different stages, including baffles~\cite{Carcasona2023}, post-processing~\cite{soni2020} and an adaptation of controls~\cite{soni2024,lough2021}. However, for the sensitivity increases aimed for in future detectors, even a single scattered light photon will pose problems. The technique of \textit{Tunable Coherence}~\cite{voigt2023} has been shown to work well in Michelson topologies and optical resonators~\cite{voigt2025} as a novel fundamental approach that could give some needed margin in dealing with scattered light. We demonstrate here that it is, in principle, also suitable for Sagnac topologies and could be a possible fundamental solution for backscattering.

\section{The concept of Tunable Coherence}
\begin{figure}[]
    \centering
    \includegraphics[width=\linewidth]{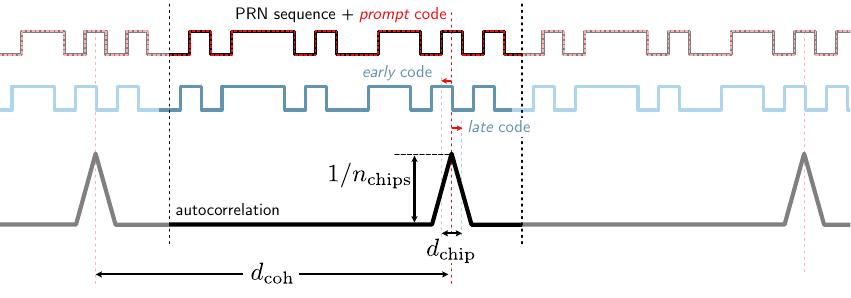}
    \caption{Sketch demonstrating the properties of the PRN-sequence to suppress scattered light, by showing the PRN-sequence and its prompt code, a shifted sequence, and the auto-correlation function. The resulting auto-correlation function between the shifted sequences only reaches 1 for matching code sequences and falls steeply for non-matching sequences reaching a value of $1/n_{\text{chips}}$ as its minimum value for optimal suppression.}
    \label{fig:prn_sequence}
\end{figure}

Tunable Coherence uses high speed phase-modulation using a pseudo-random-noise (PRN) sequence to carefully control the coherence of a laser. The imprinted phase modulation breaks the coherence and thus suppresses interference exceeding a certain relative path difference, effectively creating a pseudo-white-light interferometer~\cite{schnupp1985, voigt2023}. This minimal relative delay is given by the length of one of the chips that make up the binary modulation sequence~\cite{voigt2023}. Its corresponding optical path length is $c/f_{\text{PRN}}$, with $c$ as the speed of light and $f_{\text{PRN}}$ as the used modulation frequency. After this distance, the PRN-modulation on both beams is delayed relative to each other such that the auto-correlation function and, thus, interference becomes minimal. The auto-correlation function therefore also gives the theoretical limit of the scattered light suppression through this technique, which gives for the here chosen m-sequences a maximum suppression of $1/n_{\text{chips}}$ with $n_{\text{chips}}$ the number of chips in the sequence~\cite{sarwate1980}. As seen in Figure~\ref{fig:prn_sequence}, this optimal suppression is given, when the two interfering sequences are delayed by at least one chip relative to each other, while there is no suppression when the sequences are perfectly matched. If the sequences are delayed by only a part of a single chip, only a partial suppression is reached, depending on the length by which the sequences are delayed. 

From this follows a new relevant length, the repetition length of the used PRN-sequence, given by the number of chips $n_{\text{chips}}$ in the sequence and their length $d_{chip}$. This so called recoherence-length $d_{\text{coh}}$ thus equals
\begin{equation}
    d_{\text{coh}} = n_{\text{chips}} d_{chip} = n_{\text{chips}} c/f_{\text{PRN}}.
\end{equation}
As the used PRN-sequences repeat itself, coherence is restored for every integer multiple of this length. In any given optical setup this creates the requirement to match the delays of beams that should create interference, either in e.g. the two arms of a Michelson interferometer or in the length between a specific beam and a local oscillator as in the here shown Sagnac interferometer. For optical cavities, this corresponds to a matching of their round-trip delay to the recoherence length.

In contrast to previous usage of PRN-sequences as input for phase-modulation like digitally enhanced interferometry~\cite{shaddock2007}, where the demodulation takes place in digital post processing, here it is done optically by interfering delay-matched beams. Using a photodetector with bandwidth below the repetition frequency and then averaging over at least one full sequence per sample removes the PRN-modulation from the signal~\cite{sibley2020}. This results in the repetition rate giving a constraint for the measurement bandwidth, in order to still be able to measure the expected signals after averaging.

\section{Experimental realization}
In contrast to a Michelson topology, where one has to match both arms in length for Tunable Coherence to work, in a Sagnac topology this is inherently given as clockwise $\tau_{+\text{c}}$ and counter-clockwise $\tau_{-\text{c}}$ propagating beams are traveling the same path in opposite directions. However, as therefore the asymmetric port is also inherently dark, for the readout a balanced-homodyne-detector (BHD) is used such that now the local oscillator (LO) needs to be delay-matched. One way is to use the symmetric port as LO which would have the benefit of having traveled nearly the same total path as the asymmetric output. However, this could allow scattered light to interfere with itself traveling via the two matched paths into the readout respectively. The more reliable approach therefore is to pick off the light for the LO before the interferometer.

\subsection{Setup}
For implementing Tunable Coherence, the used laser was modulated by an electro-optical modulator (EOM) following maximum-length sequences (m-sequences) with a length between 7 and 2047 chips at $f_{\text{PRN}} = \SI{1}{\giga\hertz}$ and a modulation depth of $\pi$. 

The setup used for the tabletop experiment then consisted of a Sagnac interferometer with a circumference of around \SI{343}{\centi\meter}. As any rotational effects picked up by the interferometer would have been outside of the measurement bandwidth, a regular Sagnac instead of a zero-area was used for simplicity. 

One of the mirrors inside the interferometer was piezo-actuated to inject a simulated signal into the interferometer at $f_{\text{gw}}$. To create scattered light noise, light was picked off the counter-clockwise $\tau_{-\text{c}}$ propagating beam with a low power-reflectivity mirror ($R\approx 0.2$) and coupled into the clockwise $\tau_{+\text{c}}$ propagating beam via the same mirror. This reflectivity of $\SI{20}{\percent}$ was needed to produce enough scattered light, as it was reflected twice at this mirror. Thus $\SI{4}{\percent}$ of the incoming power were coupled back into the main beam as scattered light. This corresponds to a maximum phase error of around \SI{0.04}{\radian}. The scattered light beam was actuated by a second piezo-actuated mirror and the delay $\tau_{sc}$ relative to the interferometer beams could be adjusted with an optical delay line. In order to demonstrate the remaining coherence length introduced by Tunable Coherence, the position was chosen such that $\tau_{sc}$ could be varied between \SI{0}{\centi\meter} and \SI{50}{\centi\meter} which exceeds the theoretical estimate for \SI{1}{\giga\hertz} PRN-frequency given by $d_{\text{chip}}=c/f_{\text{PRN}}\approx\SI{30}{\centi\meter}$ with the speed of light c. The location for the injection of the simulated signal $f_{\text{gw}}$ can be chosen at random in the interferometer, as the positioning does not change the relative delay between the interfering main beams and the signal is therefore not influenced by the PRN-modulation.

\begin{figure}[]
    \centering
    \includegraphics[width=\linewidth]{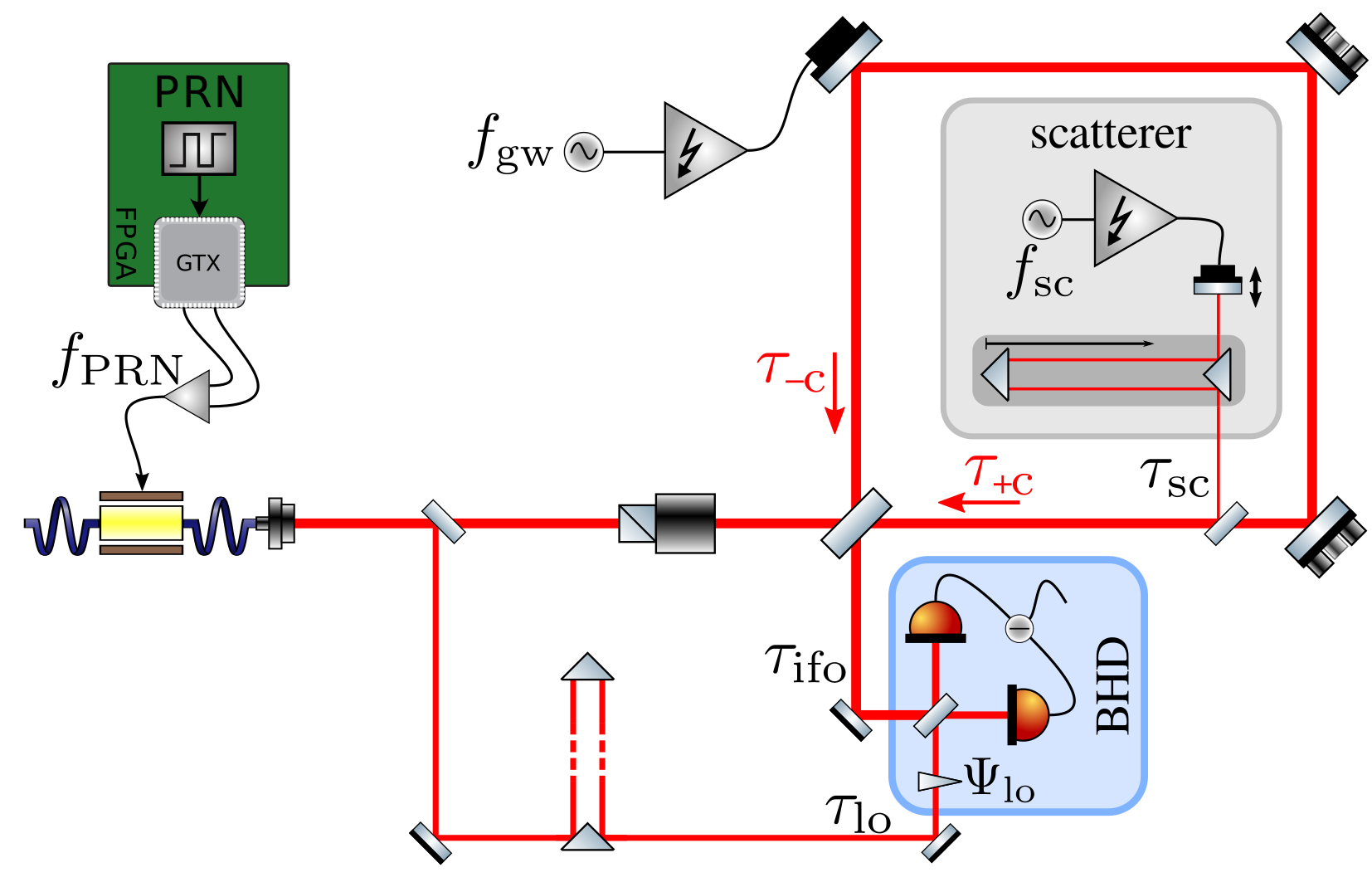}
    \caption{Sketch of the experimental setup with the laser being PRN-modulated by a fiber-coupled EOM at $f_{mod}$ and the Sagnac interferometer including the balanced homodyne readout. One of the mirrors in the interferometer was used to actuate the phase of the beams and through this inject a signal. Some light from the counter-clockwise propagating beam was reflected out and coupled back into the clockwise propagating beam after being reflected from another actuated mirror and going through an optical delay line. From this, the scattered light picks up an additional delay $\tau_{sc}$ relative to the delays $\tau_{\text{+c}}$ and $\tau_{\text{-c}}$. \componentlibrary}
    \label{fig:exp_setup}
\end{figure}

To read out the interferometer signal a BHD at the asymmetric port was used. For the local oscillator (LO) part of the light was picked off before the interferometer and delay matched to the signal beam leaving the interferometer. A mismatch in delay between the LO and the signal beam would result in a suppression of the signal from the interferometer through the PRN-modulation. 
The layout of the setup can be found in Figure~\ref{fig:exp_setup}.

The control loop for locking the LO phase with another piezo-actuated mirror had a corner frequency of around \SI{5.5}{\kilo\hertz}
such that it was effectively free-floating in the measurement range around \SI{170}{\kilo\hertz}. This range was chosen to be limited by residual laser amplitude noise only.
Two signals were injected at piezo-resonances, one with the piezo-actuated mirror directly inside the interferometer at $f_{\text{gw}}=\SI{172.4}{\kilo\hertz}$ and one to modulate the scattered light phase at $f_{\text{sc}} = \SI{170}{\kilo\hertz}$.
As these resonances were not strong enough to modulate through a full fringe, the scattered light coupling is dependent on the DC-phase due to the non-linear coupling. As this DC-phase in turn slowly fluctuates, the piezo-actuator was ramped through several fringes during the time of each measurement to guarantee that a measured timeseries always contained the strongest possible coupling. This allowed for comparison between asynchronously recorded data.

\subsection{Results}
\begin{figure*}[]
    \centering
    \includegraphics[width=\linewidth]{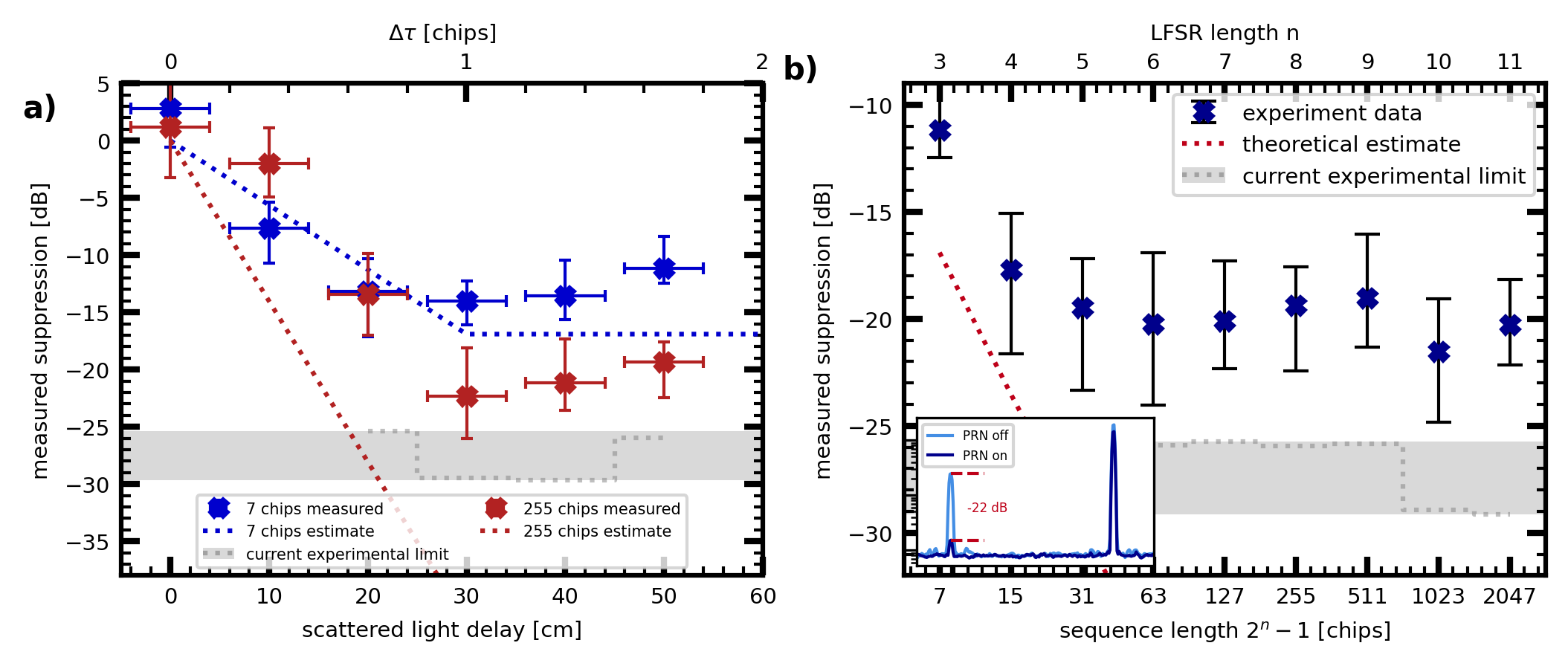}
    \caption{Measured suppression of the scattered light depending on (a) the delay of the scattered light relative to the interferometer beam and (b) the PRN-sequence length\newline
    Figure (a) shows the achieved scattered light signal suppression using 7 and 255 chip long PRN-sequences for different scattered light delays between $\SI{0}{\centi\meter}$ and $\SI{50}{\centi\meter}$ compared to the theoretical expectation and current experimental limit within the setup.
    \newline
    Figure (b) shows the achieved suppression of the scattered light signal compared to the theoretical expectation and our current experimental limit within the setup for PRN-sequence lengths between 7 and 2047 chips.
    }
    \label{fig:results}
\end{figure*}

Data was recorded in time series of two seconds
sampled at \SI{2}{\mega\hertz} for the shorter sequences (255 chips or less), as those have repetition rates that are higher than the chosen sampling rate. 
For longer sequences, starting at 511 chips, the sampling rate had to be reduced accordingly, to ensure each sample averaged over at least one full PRN-sequence. Thus time series of four and eight second were recorded respectively to keep the amount of samples for each time series constant. This also gives an upper limit for measurable frequencies when using Tunable Coherence, as the bandwidth of the measurement is limited by the sampling rate, which in turn is limited by the sequences repetition rate (typically above \SI{500}{\kilo\hertz} in this experiment). Using a bandwidth between the repetition rate and $f_\text{PRN}$ is unfeasible due to introduced artifacts~\cite{sibley2020}, while exceeding $f_\text{PRN}$ is assumed to neglect the effects of the PRN-modulation. Thus we limit our experiments to frequencies below the sequence repetition rate.

For each data point, a time series with PRN-modulation and one without it was recorded.
This data was used to compute spectra using Welch's method with a Blackman-Haris window and \SI{50}{\percent} overlap. The measured suppression was then taken as the improvement of the signal to noise ratio (SNR) between the two measurements for each data point. This SNR was calculated as the ratio between the signal peak at $f_{\text{gw}} = \SI{172.4}{\kilo\hertz}$ and the scatter peak at $f_{\text{sc}} = \SI{170}{\kilo\hertz}$.

Besides the reduction in scatter amplitude in those spectra, we also observed a small, relatively static reduction in the amplitude of the simulated signal across all measurements by $2-\SI{3}{\decibel}$. We assume this comes from a slight delay mismatch of a few \SI{}{\centi\meter} between the LO and signal field at the BHD but was not further investigated. Such an amplitude reduction was not found for previous experiments using a Michelson interferometer \cite{voigt2025}. 

In the setup, two different types of measurements were done: one where the length of the PRN-sequence was changed and one where the relative delay $\tau_{\text{sc}}$ of the scattered light was changed through the delay line positioned in the scattered light path.

For the varied relative delay we show the results in Figure~\ref{fig:results}a for the 7 and 255 chips long sequences. While the shorter sequence follows the estimate quite well and levels off around the maximum possible suppression for this sequence after exceeding the length of one chip, the longer sequence exhibits a slightly slower increase in suppression over longer delays than expected. It further levels off before reaching maximum suppression, the main reason here however is the experimental limitation for creating stronger scattered light tones.

Looking at the results for the sequence length dependence of the suppression in figure \ref{fig:results}b, the expected dependence is visible for the first three sequence lengths, before the suppression saturates. The maximal achieved suppression in this setup is $\SI{24.2}{\decibel}$ for the 1023 chips long sequence. While the early saturation is partly caused by the experimental limitations in the strength of the injected scattered light tone, all measurements clearly show some residual scattered light.

\subsection{Discussion} \label{sec:discussion}
These results show that Tunable Coherence can suppress scattered light in a simple Sagnac Speed Meter topology. However, while encouraging, the data also opens some questions. It is not fully understood why the residual scattered light cannot be fully suppressed in this setup. Limiting for the suppression are imperfections in the PRN-modulation, e.g. only if the modulation depth is exactly \SI{180}{\degree} or $\pi$, the full potential can be reached. As there is currently no control scheme implemented to guarantee this, this might be a limiting factor in our setup. We experienced the same problem in a Michelson setup, however only for suppression levels around \SI{40}{\decibel}~\cite{voigt2025}. Further, we cannot fully exclude some electronic crosstalk in the high voltage amplifier used to drive the piezos as the signal strength needed to create a significant scattered light tone was stronger than in previous setups, even though there was no crosstalk evident in the zero-light case. Another limiting factor can be the power imbalance between the interfering beams due to the scattered light only coupling into one propagation direction. This, together with the comparably small imbalance of the beam splitter, increase some readout noise sources, such as shot noise and amplitude noise. In this experiment we were, however, not yet limited by these. 
In the results for no delay between the scattered light and the main beam in Figure~\ref{fig:results}, the scattered light seems to be enhanced. This is a result of fluctuations in the strength of the scattered light peak, which dominated the measurement, as there is no suppression expected for this delay. The enhancement is therefore not physical and mostly an indicator of the non-stationarity of our experimental set-up.
Further limitations on the experiment are visible in the small reduction of the simulated signal through tunable coherence, which could be a result of a delay mismatch between the LO and the main beam of a few cm. This mismatch stemmed from the limitations of matching the lengths by hand without the use of macroscopic length tuning, which only had an accuracy of about $2-\SI{3}{\centi\meter}$ compared to the $\approx\SI{3}{\meter}$ arm-length of the Sagnac. 

While the limit in injected scattered light prohibits a clear observation of the relation between relative delay of the scattered light and achieved suppression, a trend is visible that we also observed in other topologies~\cite{voigt2025}. Especially for longer sequences a strong suppression is only reached after longer delays than estimated. As the main reason for this behavior we identify the limited bandwidth of our modulation setup, meaning the switching between chips in the PRN-sequence is not instant but has a finite rise- or fall-time, leading to a non-rectangular shape of the modulation, as observed in other experiments using GHz PRN-modulations \cite{isleif2014}. This degrades the auto-correlation-function of the sequence, limiting suppression close to the length of one chip. Generalizing this, positions at which scattered light couples back into the interferometer with a relative delay falling into this length $d_{\text{chip}}$ need to be avoided as Tunable Coherence has limited to no effect in these positions.

\section{Suppression of backscatter in ring-like topologies}

\begin{figure*}[]
    \centering
    \includegraphics[width=\linewidth]{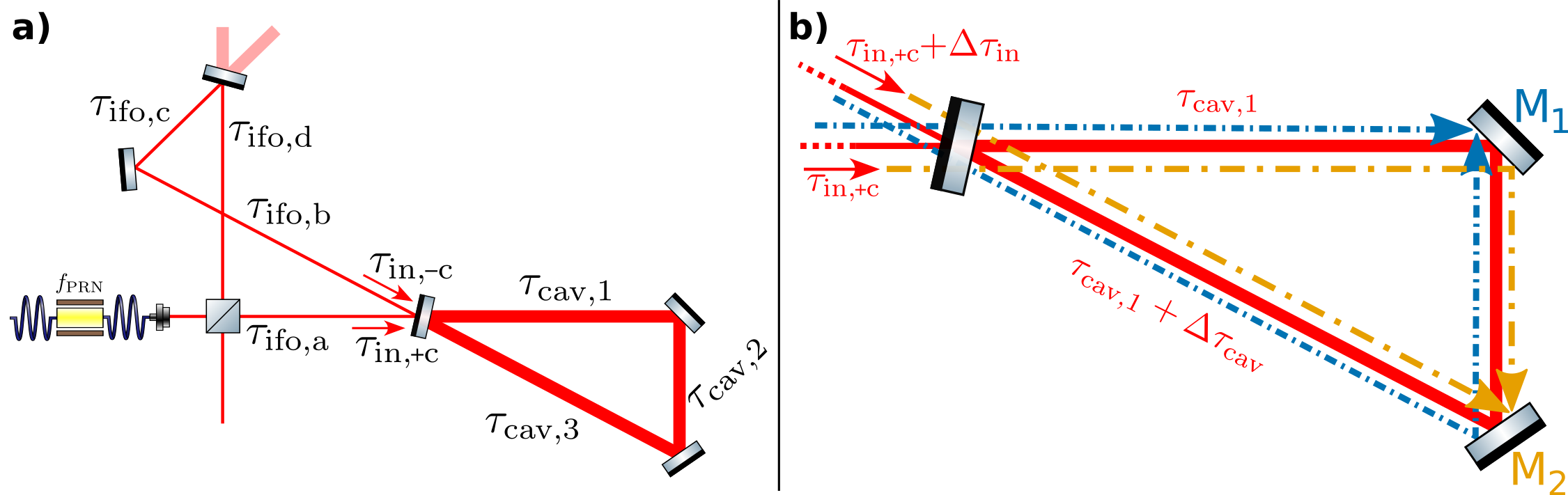}
    \caption{Sketches for the analytic backscatter discussion. In Figure (a) the designation for the variables is shown in a Sagnac Speed Meter with ring cavities. The second ring cavity is omitted, as it is analogous to the first.\\
    In (b) the needed relative delay between the two counter-propagating beams at the mirrors in a ring-cavity is visualized. The two counter-propagating beams meeting at a mirror (M1 or M2) accumulate delays before meeting, all common delays ($\tau_{\text{in,c+}}$ and $\tau_{\text{cav,1}}$) are omitted as they cancel when looking at the relative delay between the two beams. In red the new defined delays taking into account intentional mismatches in the central Sagnac and cavity. \componentlibrary}
    \label{fig:ringCavity_sketches}
\end{figure*}

Tunable coherence can also be used to suppress the light backscattering from optical surfaces directly into the counter-propagating beam. As the back-scattered light and the counter-propagating beam pick up different phases inside the interferometer, this light can be a limiting factor in a Sagnac Speed Meter introducing a power offset and scattered light noise \cite{zhang2019,pascucci2018}. This can be especially an issue in ring-cavities due to the high power and number of round-trips.

To analyze the effect of Tunable Coherence on backscatter, the delay difference between the two beams in the setup in relation to the PRN-sequence at a point of possible backscatter is used. 
Starting with the analysis of ring-cavities, a general requirement for the use of cavities with Tunable Coherence is the matching of the cavities roundtrip length to an integer multiple $n$ of the sequence's recoherence length $d_{\text{coh}}=c\cdot t_{\text{seq}}$~\cite{voigt2023, voigt2025} with $t_{\text{seq}}=n_{\text{chips}}/f_{\text{PRN}}$:

\begin{equation}
    \sum_{i}\tau_{cav,i} = n\cdot t_{\text{seq}}= n\cdot d_{\text{coh}}/c.
\end{equation}
A sketch of the setup with all variables included can be found in Figure~\ref{fig:ringCavity_sketches}a. For all delays integer multiples of the recoherence length are omitted as they would be the same in regards to Tunable Coherence.

To suppress back-scattered light, the two counter-propagating beams have to be mismatched in the delay between each other at surfaces where backscatter can occur.
For the input mirror this means
\begin{equation}
    \tau_{\text{in,+c}} \neq \tau_{\text{in,-c}}
\end{equation}
as a round trip in the cavity does not change the relative delay due to the matching of the cavity to an integer multiple of the sequence length.
From this, we get for the central Sagnac interferometer
\begin{equation}
    \tau_{\text{ifo,a}} \neq \tau_{\text{ifo,b}} + \tau_{\text{ifo,c}} + \tau_{\text{ifo,d}}
\end{equation}
or for the input in the other ring-cavity
\begin{equation}
    \tau_{\text{ifo,d}} \neq \tau_{\text{ifo,a}} + \tau_{\text{ifo,b}} +\tau_{\text{ifo,c}}
\end{equation}
From here on only one ring-cavity (shown in Figure~\ref{fig:ringCavity_sketches}b) will be considered, as the requirements for the other follow analogous. To suppress backscatter at the two mirrors inside the cavity, the following requirements have to be met:
\begin{equation}
    \tau_{\text{in,+c}} + \tau_{\text{cav,1}} \neq \tau_{\text{in,-c}} + \tau_{\text{cav,3}} + \tau_{\text{cav,2}}
\end{equation}
for the mirror M1 where the blue arrows meet in Figure~\ref{fig:ringCavity_sketches}b and
\begin{equation}
    \tau_{\text{in,-c}} + \tau_{\text{cav,3}} \neq \tau_{\text{in,+c}} + \tau_{\text{cav,1}} + \tau_{\text{cav,2}}
\end{equation}
for the mirror M2 where the orange arrows meet.
The difference in delay at the input mirror can be written as $\tau_{\text{in,-c}}-\tau_{\text{in,+c}}\,=\,\Delta\tau_{\text{in}}$ such that
\begin{align}
    \tau_{\text{cav,1}} &\neq \Delta\tau_{\text{in}} + \tau_{\text{cav,3}} +\tau_{\text{cav,2}}\\
    \Delta\tau_{\text{in}} + \tau_{\text{cav,3}} &\neq \tau_{\text{cav,1}} + \tau_{\text{cav,2}}
\end{align}
By defining the difference between the two cavity beams reaching
the input mirror as $\Delta\tau_{\text{cav}} \,= \,\tau_{\text{cav,3}} \,- \,\tau_{\text{cav,1}}$, where $\Delta\tau_{\text{cav}} \,= \,0$ is possible, this becomes
\begin{align}
    n\cdot t_{\text{seq}} &\neq \Delta\tau_{\text{in}} + \Delta\tau_{\text{cav}} + \tau_{\text{cav,2}}\\
    n\cdot t_{\text{seq}} &\neq \Delta\tau_{\text{in}} + \Delta\tau_{\text{cav}} - \tau_{\text{cav,2}}
\end{align}
with $n\in\mathbb{N}_0$ as any integer repetition of the sequence is regarded as zero relative delay. These equations are illustrated in Figure~\ref{fig:ringCavity_sketches}(b) by the blue and yellow arrows.

Finally, with the total difference in delay 
$\Delta\tau_{\text{in}}\,+\,\Delta\tau_{\text{cav}}\,=\,\Delta\tau$, the restriction comes down to
\begin{equation}
    n\cdot t_{\text{seq}} \neq \left|\Delta \tau - \tau_{\text{cav,2}} \right|.
\end{equation}
The calculations are visualized in Figure~\ref{fig:ringCavity_sketches}b in detail where for both mirrors the two counter-propagating beams are shown with their difference in relative delay. Delays both beams accumulate, like $\tau_{\text{in,+c}}$ and $\tau_{\text{cav,1}}$, are omitted.

This shows the possibility to achieve a mismatch in delay at the cavity mirrors, which in turn shows backscatter can in principle be suppressed with Tunable Coherence due to its capability to suppress interference with unequal relative delays. For this to be possible the distance between the two mirrors in the cavity cannot compensate the intended mismatches between the cavity mirrors and inside the central interferometer.
Using the same approach, the restriction on ring-cavities with more mirrors could be found, as well as the restriction on mirror placement in the central Sagnac interferometer by using a relative input delay of $\Delta\tau_{\text{in}}\,=\,0$ and the delays $\tau_{\text{ifo,i}}$.
The analysis can be extended further to other ring resonator configurations and include also optics and reflections external to the actual resonators, for example in the readout path. 

Like regular scattered light, backscatter faces strongly reduced coherence for delay mismatches exceeding the chip-length $d_{\text{chip}}$ of the PRN-modulation and is thus suppressed. While this will also apply to power fluctuations stemming from interference of backscattered light, it cannot tackle the power imbalance due to loss of power in one propagation direction, as only the phase noise is suppressed.

\section{Conclusion}
The results shown in Figure~\ref{fig:results}a and~\ref{fig:results}b are promising for using Tunable Coherence to suppress scattered light in Sagnac interferometers, showing about one order of magnitude in suppression, despite being limited by the experimental setup. One of the limitations is the complexity of the setup, which is higher than for a simple Michelson interferometer, and the resulting issues in perfectly matching the delay of the local oscillator to the signal beam coming from the interferometer.
The effectiveness of Tunable Coherence was successfully demonstrated in a Michelson topology before~\cite{voigt2025}. As we now also present promising results in a Sagnac topology, Tunable Coherence might also be interesting for Sagnac Speed Meters, as well as sloshing or EPR speed meter topolgies, which could be possible alternatives for currently Michelson based gravitational wave detectors~\cite{purdue2002,knyazev2018,danilishin2019}. 
According to an analytical inspection, Tunable Coherence also seems able to reduce the influence of phase noise due to backscatter in Sagnac Speed Meters and ring-resonator. Thereby placing some effective restrictions driven by the recoherence length of the PRN-modulation on the positioning of the mirrors. As this only reduces the phase noise, there is no change to the power imbalance resulting from backscatter. Therefore Tunable Coherence could be a more fundamental solution for reducing the phase noise due to backscatter in gyroscopes~\cite{ringlaser_igel2021} and other uses of ring-resonators, instead of removing it during post-processing~\cite{hurst2014}.

\appendix
\begin{acknowledgements}
    This research was funded by the Deutsche Forschungsgemeinschaft (DFG, German Research Foundation) under Germany's Excellence Strategy---EXC 2121 ``Quantum Universe''---390833306. 
\end{acknowledgements}

\FloatBarrier

\end{document}